\documentclass[10pt, lettersize, journal]{IEEEtran}
\normalsize
\usepackage{cite}
\usepackage[cmex10]{amsmath}
\usepackage{amssymb}
\usepackage{amsthm}
\usepackage{mathrsfs}
\usepackage{bm}
\usepackage[mathscr]{eucal}
\usepackage{amssymb,amsmath,amsthm,amsfonts,latexsym}
\usepackage{amsmath,graphicx,bm,xcolor,url}
\usepackage{graphicx}
\usepackage{latexsym}
\usepackage{CJK}
\usepackage{indentfirst}
\usepackage{psfrag}
\usepackage{setspace}
\usepackage{algorithmic}
\usepackage{algorithmic, cite}
\usepackage{algorithm}
\usepackage{array}
\usepackage{mdwmath}
\usepackage{mdwtab}
\usepackage{eqparbox}
\usepackage{url}
\usepackage{epstopdf}
\usepackage{epsfig,epsf,psfrag}
\usepackage{fixltx2e}
\usepackage{verbatim}
\usepackage{textcomp}
\usepackage{psfrag} 

\usepackage{subfigure} 
\usepackage{caption}

\theoremstyle{plain}

\theoremstyle{remark}

\theoremstyle{plain}

\theoremstyle{remark}

\theoremstyle{plain}

\theoremstyle{remark}

\theoremstyle{remark}

\theoremstyle{remark}

\theoremstyle{remark}

\theoremstyle{remark}

\theoremstyle{remark}

\catcode`~=11 \def\UrlSpecials{\do\~{\kern -.15em\lower .7ex\hbox{~}\kern .04em}} \catcode`~=13

\allowdisplaybreaks[4]

\newcommand{\calC}{\mathcal{C}}

\newcommand{\calI}{\mathcal{I}}

\newcommand{\calN}{\mathcal{N}}
\newcommand{\calO}{\mathcal{O}}

\newcommand{\calU}{\mathcal{U}}

\newcommand{\ba}{\mathbf{a}}

\newcommand{\bb}{\mathbf{b}}

\newcommand{\bC}{\mathbf{C}}

\newcommand{\bE}{\mathbf{E}}
\newcommand{\boldf}{\mathbf{f}}
\newcommand{\bF}{\mathbf{F}}
\newcommand{\bg}{\mathbf{g}}
\newcommand{\bG}{\mathbf{G}}
\newcommand{\bh}{\mathbf{h}}

\newcommand{\bI}{\mathbf{I}}

\newcommand{\bS}{\mathbf{S}}

\newcommand{\bu}{\mathbf{u}}

\newcommand{\by}{\mathbf{y}}

\newcommand{\bz}{\mathbf{z}}

\newcommand{\rmb}{\mathrm{b}}

\newcommand{\rmd}{\mathrm{d}}

\newcommand{\rmh}{\mathrm{h}}
\newcommand{\rmH}{\mathrm{H}}

\newcommand{\rmP}{\mathrm{P}}

\newcommand{\rmr}{\mathrm{r}}

\newcommand{\rmt}{\mathrm{t}}
\newcommand{\rmT}{\mathrm{T}}

\newcommand{\rmz}{\mathrm{z}}

\newcommand{\bbC}{\mathbb{C}}

\newcommand{\bbE}{\mathbb{E}}

\newcommand{\bbR}{\mathbb{R}}

\DeclareMathAlphabet{\mathbsf}{OT1}{cmss}{bx}{n}
\DeclareMathAlphabet{\mathssf}{OT1}{cmss}{m}{sl}

\DeclareSymbolFont{bsfletters}{OT1}{cmss}{bx}{n}
\DeclareSymbolFont{ssfletters}{OT1}{cmss}{m}{n}
\DeclareMathSymbol{\bsfGamma}{0}{bsfletters}{'000}
\DeclareMathSymbol{\ssfGamma}{0}{ssfletters}{'000}
\DeclareMathSymbol{\bsfDelta}{0}{bsfletters}{'001}
\DeclareMathSymbol{\ssfDelta}{0}{ssfletters}{'001}
\DeclareMathSymbol{\bsfTheta}{0}{bsfletters}{'002}
\DeclareMathSymbol{\ssfTheta}{0}{ssfletters}{'002}
\DeclareMathSymbol{\bsfLambda}{0}{bsfletters}{'003}
\DeclareMathSymbol{\ssfLambda}{0}{ssfletters}{'003}
\DeclareMathSymbol{\bsfXi}{0}{bsfletters}{'004}
\DeclareMathSymbol{\ssfXi}{0}{ssfletters}{'004}
\DeclareMathSymbol{\bsfPi}{0}{bsfletters}{'005}
\DeclareMathSymbol{\ssfPi}{0}{ssfletters}{'005}
\DeclareMathSymbol{\bsfSigma}{0}{bsfletters}{'006}
\DeclareMathSymbol{\ssfSigma}{0}{ssfletters}{'006}
\DeclareMathSymbol{\bsfUpsilon}{0}{bsfletters}{'007}
\DeclareMathSymbol{\ssfUpsilon}{0}{ssfletters}{'007}
\DeclareMathSymbol{\bsfPhi}{0}{bsfletters}{'010}
\DeclareMathSymbol{\ssfPhi}{0}{ssfletters}{'010}
\DeclareMathSymbol{\bsfPsi}{0}{bsfletters}{'011}
\DeclareMathSymbol{\ssfPsi}{0}{ssfletters}{'011}
\DeclareMathSymbol{\bsfOmega}{0}{bsfletters}{'012}
\DeclareMathSymbol{\ssfOmega}{0}{ssfletters}{'012}

\newcommand{\hatbb}{\widehat{\bb}}

\newcommand{\hath}{\widehat{h}}

\newcommand{\hatbh}{\widehat{\bh}}

\newcommand{\bphi}{\bm{\phi}}

\newcommand{\bPsi}{\bm{\Psi}}

\newcommand{\bPhi}{\bm{\Phi}}

\def\norm#1{\left\| #1 \right\|}
\def\norm2#1{\left\| #1 \right\|_2}
\def\norm22#1{\left\| #1 \right\|_2^2}

\DeclareMathOperator{\diag}{diag}

\DeclareMathOperator{\tr}{tr}

\newcommand{\qednew}{\nobreak \ifvmode \relax \else
      \ifdim\lastskip<1.5em \hskip-\lastskip
      \hskip1.5em plus0em minus0.5em \fi \nobreak
      \vrule height0.75em width0.5em depth0.25em\fi}

\usepackage{float}
\usepackage{cite}
\usepackage{times,amsmath,color,amssymb,graphicx,epsfig,multirow,float,algorithm,algorithmic}
\usepackage{caption}
\usepackage{subfigure}
\usepackage[utf8]{inputenc}

\newtheorem{remark}{Remark}
\usepackage[hidelinks]{hyperref}

\begin{document}
\captionsetup{font={footnotesize}}
\title{Channel Estimation and Training Design for Active RIS Aided Wireless Communications}

\author{
	Hao Chen, Nanxi Li, Ruizhe Long, and Ying-Chang Liang, \IEEEmembership{Fellow,~IEEE}
	\thanks{This work was supported in part by the National Key Research and Development Program of China under Grant 2018YFB1801105; in part by the Key Areas of Research and Development Program of Guangdong Province, China, under Grant 2018B010114001; in part by the Fundamental Research Funds for the Central Universities under Grant ZYGX2019Z022; and in part by the Program of Introducing Talents of Discipline to Universities under Grant B20064. \emph{(Corresponding author: Ying-Chang Liang.)}}
	\thanks{H.~Chen is with the National Key Laboratory of Wireless Communications, University of Electronic Science and Technology of China, Chengdu 611731, China, and also with the Yangtze Delta Region Institute (Huzhou), University of Electronic Science and Technology of China, Huzhou 313001, China (e-mail: {hhhaochen@std.uestc.edu.cn}).}
	\thanks{N. Li is with China Telecom Research Institute, Beijing 102209, China (e-mail: linanxi@chinatelecom.cn).}
	\thanks{R.~Long is with the National Key Laboratory of Wireless Communications, University of Electronic Science and Technology of China, Chengdu 611731, China (e-mail: {ruizhelong@gmail.com}).}
	\thanks{Y.-C.~Liang is with the Center for Intelligent Networking and Communications (CINC), University of Electronic Science and Technology of China, Chengdu 611731, China, and also with the Yangtze Delta Region Institute (Huzhou), University of Electronic Science and Technology of China, Huzhou 313001, China (e-mail:{liangyc@ieee.org}).}
}

\maketitle

\IEEEpubid{
	\begin{minipage}{\textwidth}
		\centering
		\vspace{0.8in} 
		\footnotesize
		{© 2023 IEEE. Personal use of this material is permitted. Permission from IEEE must be obtained for all other uses, in any current or future media, including reprinting/republishing this material for advertising or promotional purposes, creating new collective works, for resale or redistribution to servers or lists, or reuse of any copyrighted component of this work in other works. 
			DOI: \href{http://doi.org/10.1109/LWC.2023.3297231}{10.1109/LWC.2023.3297231}}
	\end{minipage}
}

\begin{abstract}
Active reconfigurable intelligent surface (ARIS) is a newly emerging RIS technique that leverages radio frequency (RF) reflection amplifiers to empower phase-configurable reflection elements (REs) in amplifying the incident signal. Thereby, ARIS can enhance wireless communications with the strengthened ARIS-aided links. In this letter, we propose exploiting the signal amplification capability of ARIS for channel estimation, aiming to improve the estimation precision. Nevertheless, the signal amplification inevitably introduces the thermal noise at the ARIS, which can hinder the acquisition of accurate channel state information (CSI) with conventional channel estimation methods based on passive RIS (PRIS). To address this issue, we further investigate this ARIS-specific channel estimation problem and propose a least-square (LS) based channel estimator, whose performance can be further improved with the design on ARIS reflection patterns at the channel training phase. Based on the proposed LS channel estimator, we optimize the training reflection patterns to minimize the channel estimation error variance. Extensive simulation results show that our proposed design can achieve accurate channel estimation in the presence of the ARIS noises.
\end{abstract}

\begin{IEEEkeywords}
Active reconfigurable intelligent surface (ARIS), channel estimation, reflection design, least squares.
\end{IEEEkeywords}

\section{Introduction}

Reconfigurable intelligent surface (RIS), which enables a smart wireless environment, has drawn significant attention as a promising technology for
future wireless communications\cite{LISA, RISPrinciples, RISSCIS}. In particular, RIS consists of massive cost-effective reflection elements (REs), each of which can adjust the amplitude and phase of the incident signal. 
By properly designing the reflection pattern formed with the reflection coefficients of the REs, the RIS can enhance the desired signals and/or suppress the undesired interference at the destinations, leading to the increased spectral and energy efficiency.

Currently, most existing studies focus on wireless communications aided with the passive RIS (PRIS) whose REs reconfigure the incident signal only with phase adjustments. The phase adjustments enable the PRIS to offer an outstanding performance gain when the direct links between communication nodes are blocked \cite{blockedcase}. However, for the general cases where there already exist the direct links, the performance gain provided by the PRIS is moderate, since the strength of the PRIS-aided link is much weaker than that of the direct link due to the multiplicative fading problem \cite{weakrislink}. 
Recently, active RIS (ARIS) has been proposed to overcome the severe multiplicative fading of the PRIS \cite{activeRIS}. 
In particular, the REs deployed at the ARIS are connected with not only phase shifters but also radio frequency (RF) reflection amplifiers, and therefore the ARIS is able to adjust the phases as well as to amplify the amplitude of the reflected signal. The strengthened ARIS-aided links are consequently expected for wireless communications\cite{activeRIS, samePowerBudget, SecureTransmission}. 
In \cite{samePowerBudget}, it is shown that under the same power budget, the ARIS exploits less REs but offers more achievable rate gains than the PRIS does.
In \cite{SecureTransmission}, the superiority of ARIS is also presented in secure communications, since the ARIS leads to a higher secrecy performance gain compared with the PRIS. 
Moreover, the signal amplification capability of the ARIS can also be effectively leveraged in channel estimation, which is capable to improve the estimation precision.
 
Nevertheless, while the reflection amplifiers at the ARIS can enhance the desired incoming signal, they also amplify the thermal noise generated within the ARIS itself. Consequently, the received noise comprises both the thermal noise at the receiver and the amplified noise from the ARIS, which can have a potential impact on the precision of channel estimation.
One of the critical issues to acquire the accurate CSI by channel estimation is to find a suitable training reflection pattern design for the ARIS \cite{Surveyonchannelestimation, ICASSP, OFDM, HadamardMatrix, onoffscheme}. For the conventional PRIS, the training design based on the discrete Fourier transform (DFT) matrix can be adopted for channel estimation, which can achieve the minimum estimation variance\cite{ICASSP, OFDM}. Channel estimation can also be developed based on the Hadamard matrix training design, which only requires two discrete phase-shifts for the PRIS \cite{HadamardMatrix}.
However, the conventional PRIS-based channel estimation methods and the training reflection pattern designs
cannot be directly applied to the ARIS-related channel estimation, since these designs do not take into account the amplification ability of the ARIS. Besides, the thermal noises introduced by the reflection amplifiers make it more challenging to acquire accurate CSI.

In this letter, we leverage the signal amplification ability of the ARIS to address the channel estimation problem. To do so, we propose a least-square (LS) based channel estimator and optimize the training reflection patterns of the ARIS. Specifically, the estimation error variance is minimized in the presence of the ARIS-related noises. By exploiting the prior knowledge of the ARIS-receiver channel and the structure of the DFT matrix, we simplify the problem into optimizing a single amplification factor. Moreover, we derive the estimation variance in closed form with our proposed training reflection pattern design. Simulation results demonstrate that our proposed design achieves accurate channel estimation in the presence of ARIS noises and outperforms conventional channel estimation methods.

\section{System Model}

\begin{figure}[t!]
	\centering
	\includegraphics[width=0.7\columnwidth]{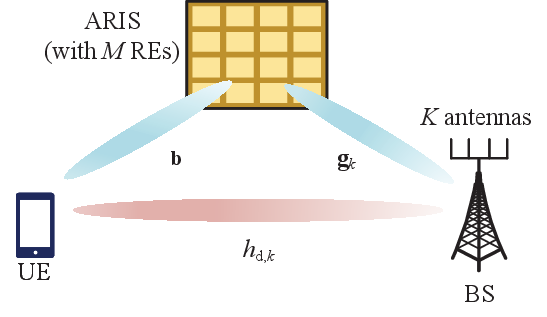}
	\vspace{-0.2cm}
	\caption{Model of an ARIS-aided system.}
	\label{systemmodel}
	\vspace{-0.5cm}
\end{figure}

As shown in Fig.~\ref{systemmodel}, we consider an ARIS-aided wireless communication system, which consists of a single-antenna UE, an ARIS with $M$ REs, and a multi-antenna BS with $K$ antennas. Time division duplex (TDD) is adopted for the considered communication system, and thus channel reciprocity is assumed such that the CSI of the downlink channel can be obtained at the BS via the uplink channel training. During the training phase, the UE transmits the pilot signals to the BS. The ARIS is configured with time-varying training reflection patterns in each training slot. Finally, the BS receives the pilot signals through the direct link and the ARIS-aided link. 
BS then can perform channel estimation based on the received pilot signals.

\subsection{Signal Model}

Denote by $\bphi_n = [\phi_{n, 1}, ..., \phi_{n, m}, ..., \phi_{n, M}]^\rmT \in \bbC^{M \times 1}$ the training reflection pattern of the ARIS in the $n$-th training slot, where $\phi_{n, m}$ is the reflection coefficient of the $m$-th RE at the ARIS. Specifically, the reflection coefficient of the $m$-th RE can be written as $\phi_{n, m} = a_{n, m} e ^ {j\theta_{n, m}}$, where $a_{n, m}$ and $\theta_{n, m}$ are the amplitude and the phase of $\phi_{n, m}$, respectively. In addition to the phase shifter, each RE in the ARIS is also integrated with an RF reflection amplifier to amplify the incident signal \cite{Dai}. Thus, the amplitude $a_{n, m}$ can be larger than $1$. We consider that the amplitude and the phase of $\phi_{n, m}$ can be continuously controlled with $a_{n, m} \in (0, a_{\rm max}]$ and $\theta_{n, m} \in (0, 2\pi]$, respectively, where $a_{\rm max} \ge 1$ is the maximum amplitude that the RF reflection amplifier can support.

The block fading channel model is assumed, which means the channels keep constant in each coherence block. Denote by $\bh_\rmd = [h_{\rmd, 1}, ..., h_{\rmd, k}, ..., h_{\rmd, K}]^\rmT \in \bbC^{K\times 1}$, $\bb \in \bbC ^ {M \times 1}$, and $\bg_k \in \bbC ^ {M \times 1}$ the direct-link channel from the UE to the BS, the forward-link channel from the UE to the ARIS, and the backward-link channel from the ARIS to the $k$-th antenna of the BS, respectively, where $h_{\rmd, k}$ is the channel response from the UE to the $k$-th antenna of the BS. Since the BS and the ARIS are generally deployed at the fixed locations with line-of-sight (LoS) channel, we assume that the backward-link channel $\bg_k, \forall k$ can be known in prior with the help of the location information at the BS. During the $n$-th training slot, the received signal at the $k$-th antenna of BS is given by
\vspace{-0.1cm}
\begin{align}\label{receivedsignal1}
    &y_k(n) \!=\! \sqrt{P_\rmT} h_{\rmd, k} s(n) \!+\! \bphi_n^\rmH \bG_k \left(\sqrt{P_\rmT} \bb s(n) \!+\! \bu(n)\right) \!+\! z_k(n) \nonumber \\
    & \!=\! \begin{bmatrix}
        1 & \bphi_n^\rmH \bG_k
    \end{bmatrix}\!
    \begin{bmatrix}
        h_{\rmd, k} \\
        \bb
    \end{bmatrix} \! \sqrt{P_\rmT} s(n) \!+\! \bphi_n^\rmH \bG_k \bu(n) \!+\! z_k(n),
\end{align}
where $P_\rmT$ is the transmit power of the UE, $s(n)$ is the pilot signal with $|s(n)| ^ 2 = 1$, $\bG_k = \diag(\bg_k)$ is the backward-link channel of the $k$-th BS antenna, $\bu(n) \sim \calC\calN(\bm{0}, \sigma_1^2 \bI_M)$ and $z_k(n)\sim \calC\calN(0, \sigma_2^2)$ are the additive white Gaussian noise (AWGN) at the ARIS and BS, respectively. 
Different from the received signal model in most studies on the conventional PRIS, 
the AWGN is also introduced at the ARIS side due to the signal amplification, which leads to an increase in channel estimation errors. 

In this model, we are interested in estimating $\bh_\rmd$ and $\bb$ with $N$ training slots. During each slot, the UE sends a pilot signal, and the ARIS configures its training reflection pattern. After receiving $N$ consecutive pilot signals, the BS estimates $\bh_\rmd$ and $\bb$.
Stacking the received signal at the $k$-th antenna across $n = 1, ..., N$ training slots, we obtain that
\vspace{-0.1cm}
\begin{align}\label{receivedsignal2}
\by_k = \sqrt{P_\rmT} \bS \bPhi_k \bh_k + \bar{\bz}_k,
\end{align}
where $\bh_k = [h_{\rmd, k}, \bb^\rmT] ^ \rmT$ is the channel vector to be estimated, $\by_k = [y_k(1), ..., y_k(N)]^\rmT$ is the received signal at the $k$-th antenna, $\bS = \diag([s(1), ..., s(N)])$ is a diagonal matrix of the pilot signals, $\bar{\bz}_k = \bPsi_k \bar{\bu} + \bz_k$ is the equivalent noise combining the AWGN at the BS and the ARIS with $\bar{\bu} = [\bu^\rmT(1), ..., \bu^\rmT(N)] ^ \rmT$ and $\bz_k = [z_k(1), ..., z_k(N)] ^ \rmT$, and the related matrices are given as
\vspace{-0.1cm}
\begin{align}
	\bPhi_k = \begin{bmatrix}
		1 & \bphi_1^\rmH \bG_k \\
		\vdots & \vdots\\
		1 & \bphi_N^\rmH \bG_k
	\end{bmatrix}, \bPsi_k = \begin{bmatrix}
		\bphi_1^\rmH \bG_k &  &  \bm{0} \\
		& \ddots & \\
		\bm{0} & & \bphi_N^\rmH \bG_k
	\end{bmatrix},
	\nonumber
\end{align}
respectively.
In (\ref{receivedsignal2}), the matrix $\sqrt{P_\rmT} \bS \bPhi_k$ is known at the BS as the observation matrix to estimate the channel vector $\bh_k$. In addition, the equivalent noise follows the distribution as $\bar{\bz}_k \sim \calC\calN(\bm{0}, \bC_{\rmz, k})$, where $\bC_{\rmz, k} = \sigma_1^2\bPsi_k\bPsi_k^\rmH + \sigma_2^2\bI_N$ is the covariance matrix of $\bar{\bz}_k$ which is a diagonal matrix but not necessarily a scaled identity matrix. 
Note that the ARIS training reflection patterns can affect the noise power of $\bar{\bz}_k$, since the reflection amplifier also amplifies the noises at the ARIS. Therefore, it is necessary to design the training reflection patterns to avoid excessive noise amplification during the channel estimation processing.

\subsection{LS-Based Estimator}
To ensure that $\bh_k$ can be uniquely estimated, it is assumed that $N\geq M+1$.\footnote{The ARIS is proved to be more efficient in improving wireless communications using much fewer REs compared to the PRIS \cite{activeRIS}. Hence, the required training overhead of $M + 1$ can be more affordable for the ARIS.} Based on this assumption and the general linear model in (\ref{receivedsignal2}), it is straightforward to find that the LS estimator is a minimum variance unbiased (MVU) estimator for $\bh_k$\cite{kay1993fundamentals}. Therefore, $\bh_k$ can be estimated as
\vspace{-0.1cm}
\begin{align}\label{estimation1}
    \hatbh_k = \left[\hath_{\rmd, k}, \hatbb_k^\rmT\right]^\rmT = \bE_k \by_k,
\end{align}
where the LS-based estimator can be written as
\vspace{-0.1cm}
\begin{align} \label{LSestimator}
    \bE_k = \frac{1}{\sqrt{P_\rmT}}  \left(\bPhi_k^\rmH \bS^\rmH \bC_{\rmz, k}^{-1} \bS \bPhi_k\right)^{-1}  \bPhi_k^\rmH \bS^\rmH \bC_{\rmz, k}^{-1}.
\end{align}
Furthermore, the covariance matrix for the estimation $\hatbh_k$ can be expressed as
\vspace{-0.1cm}
\begin{align}\label{covariancehk}
    \bC_{\rmh, k} & = \frac{1}{P_\rmT} \left(\bPhi_k^\rmH \bS^\rmH \bC_{\rmz, k}^{-1} \bS \bPhi_k\right)^{-1} \nonumber \\
    & \overset{(a)}{=} \frac{1}{P_\rmT}\left(\bPhi_k^\rmH\left(\sigma_1^2\bPsi_k\bPsi_k^\rmH + \sigma_2^2\bI_N\right)^{-1}\bPhi_k\right)^{-1},
\end{align}
where $(a)$ holds due to the fact that $\bS^\rmH\bS = \bI_N$ and $\bC_{\rmz, k}$ is a diagonal matrix. 
Moreover, the BS can perform channel estimation in (\ref{estimation1}) with each BS antenna $k$. Then, the direct-link channel can be derived as $\hatbh_\rmd = [\hath_{\rmd, 1}, ..., \hath_{\rmd, K}]^\rmT$, while the forward-link channel can be estimated with $K$ replicas. To reduce the estimation variance, the forward-link channel $\bb$ can be obtained by equally combining the $K$ replicas as
\vspace{-0.1cm}
\begin{align} \label{averaging}
    \hatbb = \frac{1}{K}\sum_{k = 1}^{K} \hatbb_k.
\end{align}
As a result, the sum of the estimation variances of $\hatbh_\rmd$ and $\hatbb$ is given as 
\vspace{-0.1cm}
\begin{align} \label{sumepsilon}
	\epsilon & = \bbE\left[\|\bh_\rmd - \hatbh_\rmd\|^2\right] + \bbE\left[\|\bb - \hatbb\|^2\right] \nonumber \\
	& = \sum_{k = 1}^{K}\left[\bC_{\rmh, k}\right]_{1, 1} + \frac{1}{K^2}\sum_{k = 1}^{K}\left(\tr\{\bC_{\rmh, k}\} - \left[\bC_{\rmh, k}\right]_{1, 1}\right),
\end{align}
where $[\bC_{\rmh, k}]_{m, m}$ denotes the $m$-th diagonal element of the matrix $\bC_{\rmh, k}$.
It is found that the estimation variance is related to the covariance matrix of $\bh_k$. Thereby optimizing the reflection patterns, a lower estimation variance can be expected.

\section{Training Reflection Pattern Design for Channel Estimation} \label{sectionIII}
In this section, the training reflection patterns are optimized to minimize the estimation variance for ARIS-aided systems.
\subsection{Problem Formulation}
The proposed LS estimator is able to simultaneously obtain $\hatbh_\rmd$ and $\hatbb$, and the sum of their estimation variances in (\ref{sumepsilon}) can be minimized by solving the following problem 
\vspace{-0.1cm}
\begin{subequations}\label{problem1}
\begin{align}
    {\rm P1:}\min_{\{\bphi_n\}\!_N} &  \sum_{k = 1}^{K}\!\left[\bC_{\rmh, k}\right]_{1, 1} \!\!+\! \frac{1}{K^2}\!\sum_{k = 1}^{K}\!\left(\!\tr\{\bC_{\rmh, k}\} \!-\! \left[\bC_{\rmh, k}\right]_{1, 1}\!\right) \label{objective1} \\
    {\rm s.t.} \enspace & \left|\phi_{n, m}\right| \le a_{\rm max}, \forall n, m. \label{constraint1}
\end{align}
\end{subequations}
Specifically, (\ref{constraint1}) is the amplitude constraint for each RE of the ARIS. 
It is challenging to solve ${\rm P1}$ with the optimal solution because of the non-convex objective function (\ref{objective1}). 
For the sake of finding a feasible reflection pattern design, we leverage a unique matrix structure that enables the product of each training pattern and the backward-link channel to be part of each row of the scaled DFT matrix.
This approach allows for a more efficient and practical training process, resulting in a more effective design for the training reflection patterns.

\subsection{Training Reflection Pattern Design}
In order to show the main idea of the proposed solution, we begin with the case that the BS is equipped with one single antenna, i.e., $K=1$. Therefore, the subscript $k$ is omitted for the single-antenna case.

\subsubsection{Single-Antenna Case}
For the single-antenna case, the objective of the optimization problem ${\rm P1}$ can be degraded into the following
\vspace{-0.1cm}
\begin{subequations}
\begin{align}
	{\rm P2:}\min_{\{\bphi_n\}_N} \enspace & \tr\{\bC_{\rmh}\} \label{objective2}\\
	{\rm s.t.}  \enspace & {\rm (\ref{constraint1})}. \label{constraint2}
\end{align}
\end{subequations}
Thanks to the fact that the LS estimator is the MVU estimator for the general linear model in (\ref{receivedsignal2}), it can attain the CRLB, i.e., $[\bC_\rmh]_{m, m} = [\calI^{-1}(\bh)]_{m, m}$, where $\calI(\bh)$ is the Fisher information matrix of $\bh$. Meanwhile, there exists a lower bound on the CRLB for each element as follows \cite{kay1993fundamentals}
\vspace{-0.1cm}
\begin{align} \label{lowerbound1}
	\left[\calI^{-1}\left(\bh\right)\right]_{m, m} \ge \frac{1}{\left[\calI\left(\bh\right)\right]_{m, m}}, m = 1, 2, ..., M + 1,
\end{align}
where the equality holds with $\calI^{-1}(\bh)$ being a diagonal matrix. This indicates that the objective (\ref{objective2}), which is also the CRLB of $\bh$, can reach its lower bound with a training reflection pattern design that enables $\bC_\rmh$ to be a diagonal matrix.

The DFT matrix-based training reflection pattern design enables the estimation covariance matrix to be a diagonal matrix for the PRIS\cite{ICASSP}. 
Inspired by that, we exploit the structure of the DFT matrix to simplify problem $\rmP2$. 
Specifically, denote the first $M + 1$ columns of the $N$-point DFT matrix by $\bF_{N, M + 1}$, and the last $M$ columns of the matrix $\bF_{N, M + 1}$ by $\bF_{N, M}$,
the rows of which are orthogonal to each other. By letting $\bphi_n^\rmH\bG$ being the scaled $n$-th row of $\bF_{N, M}$, we give the structure of $\{\bphi_n\}_N$ as follows
\vspace{-0.1cm} 
\begin{align} \label{rcdesign}
	\left[\bG^\rmH\bphi_1, \cdots, \bG^\rmH\bphi_N\right] ^ \rmH = \beta \bF_{N, M},
\end{align}
where $\beta\!\in\!\bbR^+$ is a positive real scaling factor that is introduced to adjust the signal amplification at the ARIS. Thereby, each training reflection pattern is derived as
\vspace{-0.1cm}
\begin{align} \label{reflectionpattern}
	\bphi_n = \beta \left(\boldf_n^\rmH \bG^{-1}\right) ^ \rmH, n = 1, 2, ..., N,
\end{align}
where $\boldf_n^\rmH$ is the $n$-th row of $\bF_{N, M}$. 
Based on that, we have $\bPsi \bPsi^\rmH = M \beta^2\bI_N$, which leads to the covariance matrix of the equivalent noise $\bC_\rmz$ being a scaled identity matrix. Furthermore, with $\bPhi^\rmH\bPhi = N \diag([1, \beta^2, ..., \beta^2])$, the covariance matrix for the estimation $\hatbh$ is derived as 
\vspace{-0.1cm}
\begin{align}\label{Chk}
    \bC_\rmh = \frac{M \sigma_1^2 \beta^2 + \sigma_2^2}{P_\rmT}\left(\bPhi^\rmH\bPhi\right)^{-1}.
\end{align}
Since $(\bPhi^\rmH\bPhi)^{-1}$ is a diagonal matrix, the lower bound of (\ref{objective2}) can be given as
\vspace{-0.1cm}
\begin{align}
    \epsilon^{\rm lb}\left(\beta\right) = \frac{1}{P_\rmT N}\left(M\left(\sigma_1^2\beta^2 + \sigma_2^2\beta^{-2}\right) + M^2 \sigma_1^2 + \sigma_2^2\right).
\end{align}
Therefore, the only parameter that needs to be optimized in the reflection pattern design is the scaling factor $\beta$. Problem ${\rm P2}$ can be rewritten as follows
\vspace{-0.1cm}
\begin{subequations}\label{problem2}
	\begin{align}
		{\rm P2-1:}\min_{\beta\in \bbR^+} \enspace & \epsilon^{\rm lb}\left(\beta\right) \label{objective3} \\
		{\rm s.t.}  \enspace & {\rm (\ref{constraint1}), (\ref{reflectionpattern})}.
	\end{align}
\end{subequations}
Note ${\rm P2 - 1}$ is the optimization of one single scalar with the convex objective function (\ref{objective3}). Since the LoS channel model is considered for the backward-link channel, we can obtain the range of $\beta$ by substituting (\ref{reflectionpattern}) into (\ref{constraint1}), which is obtained as 
\begin{align} \label{alpha_range}
	0 < \beta \le a_{\rm max} \sqrt{\rho_g},
\end{align}
where $\rho_g$ is the large-scale fading coefficient of the backward-link channel. Moreover, by setting $\partial \epsilon^{\rm lb}(\beta) / \partial \beta$ to $0$, the optimal value of $\beta$ can be derived as
\begin{align}\label{alpha}
	\beta_{\rm opt} = \min\left\{a_{\rm max} \sqrt{\rho_g}, \sqrt{\tfrac{\sigma_2}{\sigma_1}}\right\}.
\end{align}
With the substitution of the optimal $\beta_{\rm opt}$ into (\ref{reflectionpattern}), the training reflection pattern design $\{\bphi_n\}_N$ is attained for the ARIS.

\subsubsection{Multi-Antenna Case}
With the solution to the single-antenna case, we extend the results to the general multi-antenna case. 
In particular, the training reflection patterns $\{\bphi_n\}_N$ can be firstly designed as in (\ref{reflectionpattern}) with $\bG$ replaced by the backward-link channel related to the first BS antenna $\bG_1$.
Regarding the other BS antennas, we notice there only exists an overall phase difference between $\bphi_n^\rmH \bG_1$ and $\bphi_n^\rmH \bG_k$, i.e., $\bphi_n^\rmH \bG_1 = e^{j\Delta \theta_k} \bphi_n^\rmH \bG_{k}$, where $\Delta \theta_k$ is the overall phase difference. Therefore, the covariance matrices as the diagonal ones in (\ref{Chk}) can be held for the other BS antennas, which are free of the impact of the overall phase differences. 
The covariance matrices of the estimation $\hatbh_k$ is obtained as in (\ref{Chk}) with $\bPhi_k^\rmH \bPhi_k = N \diag([1, \beta^2, ..., \beta^2])$. 
With the substitution of $\bC_{\rmh, k}, \forall k$ into (\ref{objective1}), the sum of the estimation variances can be derived as
\vspace{-0.1cm}
\begin{align}
	\epsilon\!\left(\beta\right) \!=\! \frac{1}{P_\rmT N}\!\left(\!M\!\left(\!K\sigma_1^2\beta^2 \!+\! \frac{\sigma_2^2}{K}\beta^{-2}\!\right) \!+\! \frac{M^2\sigma_1^2}{K} \!+\! K \sigma_2^2\!\right)\!.
\end{align}
Similar to ${\rm P2}$ for the single-antenna case, ${\rm P1}$ can also be simplified as the optimization of $\beta$ to minimize $\epsilon(\beta)$ with the constraint of $\beta$ in (\ref{alpha_range}). By setting $\partial \epsilon(\beta) / \partial \beta$ to $0$, we can derive the optimal $\beta$ as
\vspace{-0.1cm}
\begin{align} \label{optalpha}
	\beta_{\rm opt} = \min \left\{a_{\rm max}\sqrt{\rho_g}, \sqrt{\tfrac{\sigma_2}{K\sigma_1}}\right\}.
\end{align}
Finally, the training reflection patterns $\{\bphi_n\}_N$ are derived by the substitution of (\ref{optalpha}) into (\ref{reflectionpattern}) with the knowledge of $\bG_1$. In particular, the training reflection pattern design requires the knowledge of $a_{\rm max}$, $\rho_g$, $K$, $\sigma_1^2$ and $\sigma_2^2$, which can be known or precisely measured as the prior knowledge.

Meanwhile, the estimation in (\ref{estimation1}) and (\ref{LSestimator}) can be reduced as
\begin{align} \label{estimation2}
	\hatbh_k = \frac{1}{\sqrt{P_\rmT}}  \left(\bPhi_k^\rmH \bPhi_k\right)^{-1}  \bPhi_k^\rmH \bS^\rmH \by_k,
\end{align}
where the inverse matrix can be easily derived with a closed form as $(\bPhi_k^\rmH \bPhi_k)^{-1} = \frac{1}{N}\diag([1, 1/\beta_{\rm opt}^2, ..., 1/\beta_{\rm opt}^2])$. Due to the matrix multiplication with each BS antenna, the arithmetic cost of our proposed method is derived as $\calO(KN\log N)$ \cite{ICASSP}.

Furthermore, the estimation variances for the direct-link channel and the forward-link channel on each element can be respectively derived as
\vspace{-0.1cm}
\begin{subequations}\label{epsilondb}
	\begin{align} 
		&\epsilon_\rmd = \frac{1}{K}\bbE\left[\|\bh_\rmd - \hatbh_\rmd\|^2\right] = \frac{\left(M\sigma_1^2\beta_{\rm opt}^2 + \sigma_2^2\right)}{P_\rmT N}, \\
		&\epsilon_\rmb = \frac{1}{M} \bbE\left[\|\bb - \hatbb\|^2\right] = \frac{\left(M \sigma_1^2 + \sigma_2^2\beta_{\rm opt}^{-2}\right)}{P_\rmT KN }.
	\end{align}
\end{subequations}
Note that the estimation variances can be decreased with more transmit power $P_\rmT$ or a larger number of pilot signals $N$.
\begin{remark}
	In (\ref{epsilondb}), we notice the estimation variance for the direct-link channel $\epsilon_\rmd$ increases as a function of $\beta_{\rm opt}$, while that for the forward-link channel $\epsilon_\rmb$ decreases by contrary. This is because the estimation of $\bh_\rmd$ is irrelevant to the training reflection patterns of the ARIS. With larger $\beta_{\rm opt}$ and thus more signal amplification at the ARIS, the power of the received equivalent noise turns larger, which results in the worse performance of the estimation of $\bh_\rmd$. As for the forward-link channel, the estimation of $\bb$ is dependent on the ARIS. Even though the received noise is enlarged with larger $\beta_{\rm opt}$, the received signal through the ARIS-aided link is amplified more significantly, which leads to better performance of the estimation of $\bb$.
\end{remark}

Based on the optimized training reflection patterns, channel estimation based on minimum mean square error (MMSE) can also be derived with the prior known statistics of both the channel and noise. Moreover, in scenarios where the backward link is not fixed and slow-varying, 
the BS can design training reflection patterns with the backward-link CSI which is estimated over a long period, and then feed the designed patterns to the ARIS. Thereby, within a shorter period, the fast-varying direct- and forward-link channels can be estimated by our proposed scheme.

\section{Simulation Results}

In this section, we present the simulation results that demonstrate the effectiveness of our proposed channel estimator, which incorporates the proposed design for training reflection patterns. To evaluate the performance of our approach, we conducted simulations using square-shaped, half-wavelength-spaced uniform planar arrays (UPAs) on both the ARIS and the BS.
The number of the REs and the number of the BS antennas are set as $M = 16$ and $K = 16$, respectively. The backward-link channel is modeled as $[\bg_1, ..., \bg_K] = \sqrt{\rho_g} \ba(\theta_\rmr, \varphi_\rmr) \ba^\rmH(\theta_\rmt, \varphi_\rmt)$, where $\ba(\theta, \varphi)$ is the steering vector dependent on the direction of $(\theta, \varphi)$ with $\theta_\rmr,\theta_\rmt \sim \calU(0, \pi)$ and $\varphi_\rmr,\varphi_\rmt \sim \calU (-\frac{\pi}{2}, \frac{\pi}{2})$ being the horizontal and vertical angles of the direction of arrival and departure, respectively. The large-scale fading coefficient is set as $\rho_g = 10^{-3}d^{-2}$, where the distance between the ARIS and the BS is set as $d = 50$ m. The direct-link and forward-link channels are modeled as Rayleigh fading channels. The AWGN power at the BS and ARIS are set as $\sigma_1^2 \!=\! -70$ dBm and  $\sigma_2^2 \!=\! -80$ dBm, respectively. The conventional DFT-based scheme \cite{ICASSP, OFDM} and the on-off scheme \cite{onoffscheme} are considered as the benchmark schemes. In particular, the forward-link channel is derived by the removal of the prior known backward-link channel from the estimated cascaded-link channel. The results of mean square errors (MSEs) are derived by $10^4$ independent trials.

\begin{figure}[t!]
	\centering
	\includegraphics[width=0.65\columnwidth]{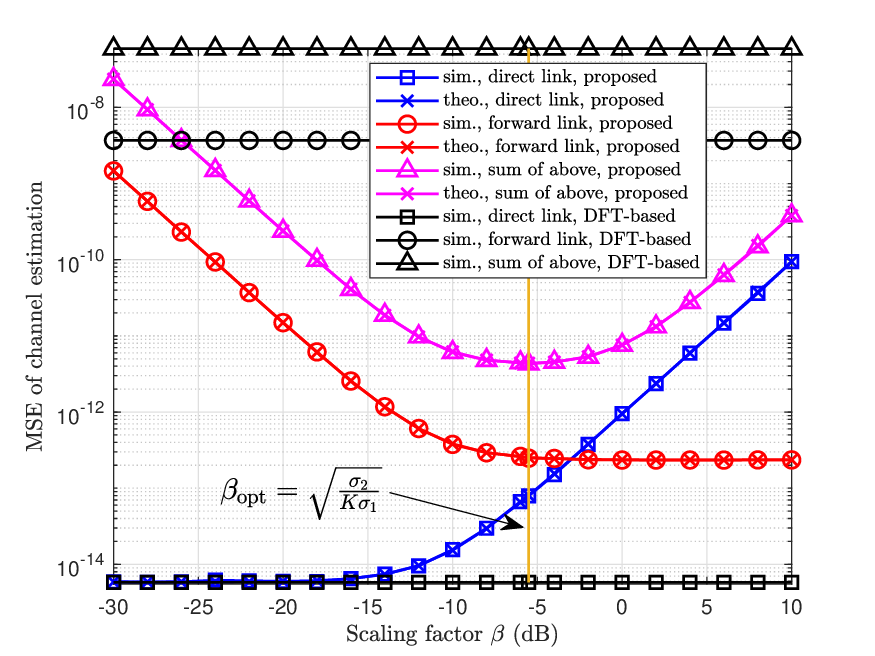}
	\vspace{-0.2cm}
	\caption{MSE of channel estimation versus scaling factor $\beta$.}
	\vspace{-0.5cm}
	\label{MSE_vs_alpha}
\end{figure}
Fig.~\ref{MSE_vs_alpha} depicts the MSEs of channel estimation versus the scaling factor $\beta$. The maximum amplitude $a_{\rm max}$ is assumed to be sufficiently large, and thus $\beta_{\rm opt} \!=\! \sqrt{\sigma_2/K\sigma_1}$. Since the conventional DFT-based scheme has no optimization on the scaling factor, its performance keeps constant with the increase in $\beta$. First, we find that the theoretical and simulated results overlap each other with our proposed scheme. Moreover, our proposed scheme outperforms the benchmark scheme significantly in terms of the sum of the MSEs. When $\beta = \beta_{\rm opt}$, the sum of the MSEs reaches its minimum value, which is consistent with our analysis. Additionally, we observe that the MSE of the estimated forward-link channel by our proposed scheme decreases with the increase in $\beta$. When $\beta \ge -5$ dB, it converges to a value which is much lower than the MSE by the benchmark scheme. On the contrary, with the increase in $\beta$, the MSE of the estimated direct-link channel by our proposed scheme increases, which is higher than that of the benchmark scheme. In other words, our proposed scheme can significantly improve the performance of channel estimation for the forward-link channel, but worsen that for the direct-link channel. Overall, compared with the benchmark scheme, our proposed scheme can promote the performance of channel estimation with the optimal value of $\beta$.

\begin{figure}[t!]
	\centering
	\subfigure[Sum of MSEs versus $P_\rmT$]{
	\begin{minipage}{0.47\columnwidth}
		\centering
		\includegraphics[width=1\columnwidth]{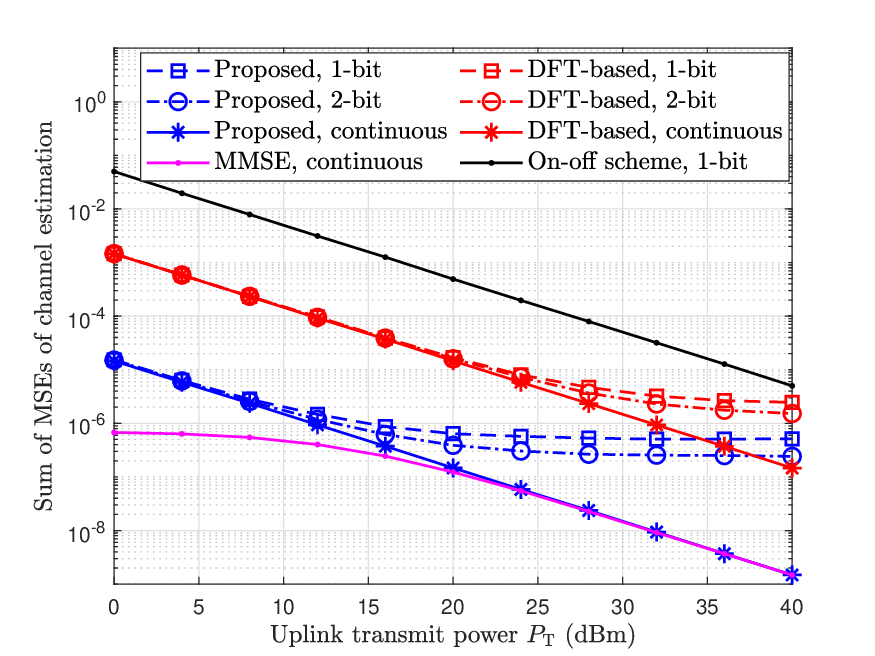}
		\vspace{-0.2cm}
		\label{MSE_vs_power}
	\end{minipage}}
	\subfigure[Achievable rate versus $P_\rmT$]{
	\begin{minipage}{0.47\columnwidth}
		\centering
		\includegraphics[width=1\columnwidth]{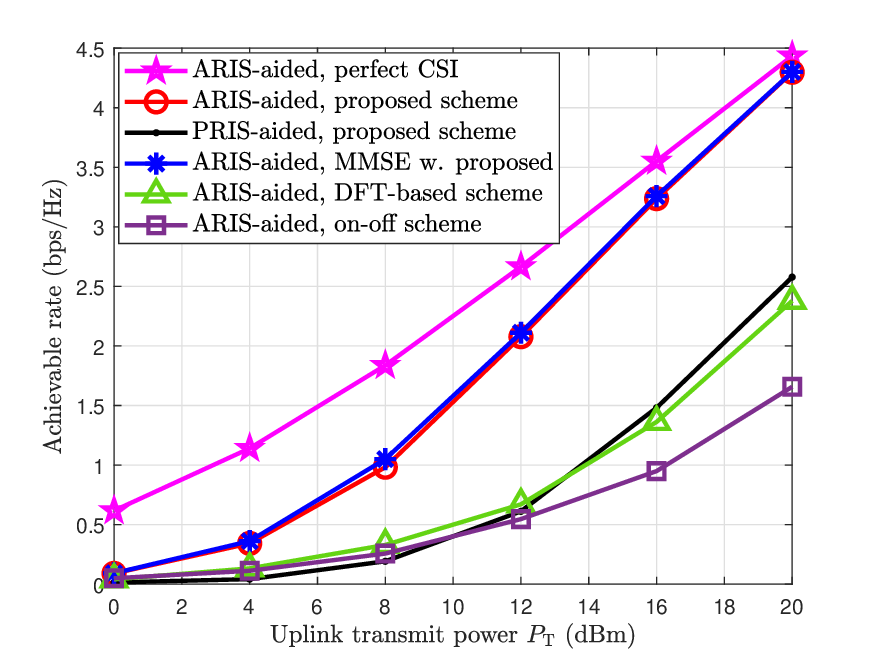}
		\vspace{-0.2cm}
		\label{rate_vs_power}
	\end{minipage}}
	\vspace{-0.2cm}
	\caption{Performance versus transmit power $P_\rmT$, where $a_{\rm max}^2 \!=\! 20$ dB.}
	\vspace{-0.5cm}
\end{figure}
Fig.~\ref{MSE_vs_power} depicts the sum of MSEs versus uplink transmit power $P_\rmT$ under different discrete ARIS phase shifts. With our proposed training design, the MMSE estimation can be derived as the lower bound for channel estimation. First, we observe that our proposed scheme outperforms the two benchmark schemes under the same discrete phase shift. For the case with continuous phase shifts, our proposed method, which is based on LS, achieves almost identical performance to that of MMSE estimation when $P_\rmT \ge 20$ dBm. Moreover, with 2-bit discrete phase shifts, the MSE curve by our proposed scheme levels off once $P_\rmT$ exceeds $28$ dBm. This is because the orthogonality among the rows of $\bPhi_k$ is partially disrupted by discrete phase shifts, which leads to a decline in channel estimation.

Fig.~\ref{rate_vs_power} depicts the achievable rate versus uplink transmit power $P_\rmT$. With our proposed channel estimation, the ARIS-aided system is compared with the PRIS-aided system where $a_{\rm max}^2 \!=\! 0$ dB and no thermal noise is introduced at the PRIS. With perfect and estimated CSI, the joint beamforming vectors are optimized as \cite{activeRIS}. The results of achievable rate are derived with the optimized beamforming vectors. First, the results show that the ARIS-aided system outperforms the PRIS-aided system thanks to the signal amplification of the ARIS. Moreover, since our proposed scheme provides more precise channel estimation, the curves by our proposed scheme are closer to the curve obtained with perfect CSI than those generated by the benchmark schemes.

\begin{figure}[t!]
	\centering
	\includegraphics[width=0.65\columnwidth]{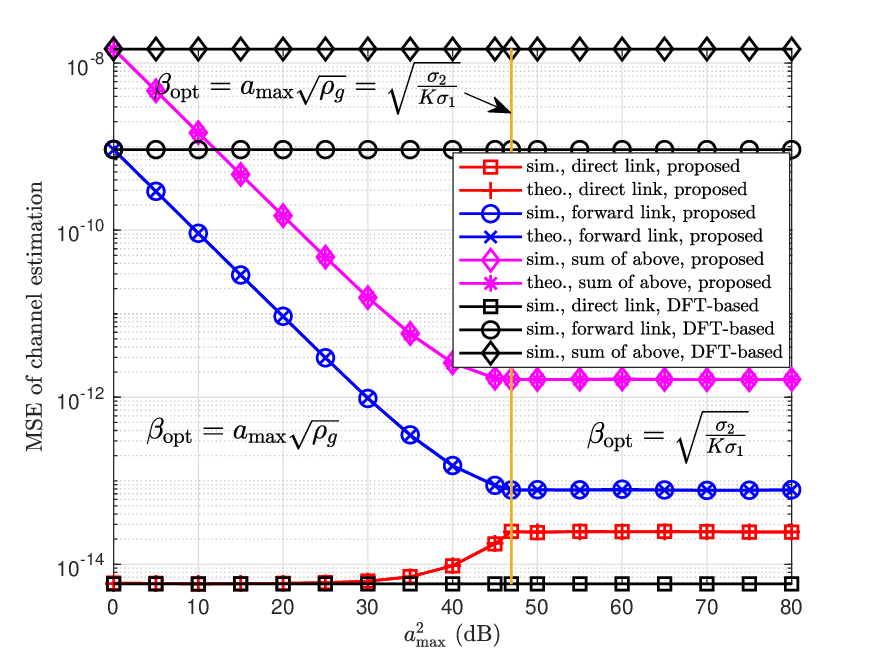}
	\vspace{-0.2cm}
	\caption{MSE of channel estimation versus $a_{\rm max}^2$.}
	\vspace{-0.5cm}
	\label{MSE_vs_amax}
\end{figure}

Fig.~\ref{MSE_vs_amax} depicts the MSE of channel estimation versus the maximum amplitude of each reflection coefficient of the ARIS $a_{\rm max}^2$. 
Within the range of investigation, it is found that the MSE for both the sum and the forward-link channel decreased as the maximum amplitude $a_{\rm max} ^2$ increasing, while the MSE for the direct-link channel increases. This trend continues until the maximum amplitude reaches a value of $a_{\rm max}^2 = 48$ dB, above which all of the MSE curves plateau. This indicates that the optimal scaling factor is no longer dependent on the maximum amplitude $a_{\rm max}^2$.
This means that $\beta_{\rm opt}$ achieves the performance balance between the direct-link channel and the forward-link channel with $a_{\rm max}^2 \!=\! 52$ dB.

\section{Conclusions}

In this letter, we propose a LS-based channel estimator and a training reflection pattern design for ARIS-aided wireless communications. The direct- and forward-link channels are estimated with the proposed estimator. To minimize the estimation variance, the training reflection patterns of the ARIS are optimized under the constraint of each reflection coefficient of the ARIS. The solution can be derived based on the DFT matrix and the optimized scaling factor with the prior knowledge of the ARIS-receiver channel. Simulation results are provided to validate that our proposed scheme can outperform the conventional one with more accurate channel estimation.

\bibliography{IEEEabrv,reference}

\bibliographystyle{IEEEtran}

\end{document}